\documentclass[twocolumn,aps,pra]{revtex4-1}%,linenumbers
\usepackage{epsfig}
\usepackage[english]{babel}
\usepackage{latexsym}
\usepackage{subfigure}
\usepackage{graphics}
\usepackage{epstopdf}
\usepackage{dcolumn}
\usepackage{amsmath}
\usepackage{hyperref}
\usepackage{amssymb}
\usepackage{appendix}
\usepackage{color}
\usepackage{lineno}
\usepackage{soul}
\usepackage{booktabs}
\usepackage{longtable}
\usepackage{multirow}
\usepackage{array}
\usepackage{ulem}
\begin{document}

\title{Caustic effects on high-order harmonic generation in graphene}
%\title{Caustic singularity  induced harmonic enhancement structure of graphene}

\author{Fulong Dong$^{1}$, Qinzhi Xia$^{2}$, Jie Liu$^{1,*}$}

\date{\today}

\begin{abstract}
We employ the two-band density-matrix equations and time-dependent density functional theory to calculate high-order harmonic generation (HOHG) in graphene under a femtosecond laser irradiation.
Our investigation uncovers a striking harmonic enhancement structure (HES) within a specific energy range of the HOHG spectrum.
In this regime, we find the convergence of multiple interband electron-hole recombination trajectories, leading to the zero determinant of the Hessian matrix of the semiclassical action.
This trajectory convergence exhibits the characteristics akin to the focusing behavior of light rays, commonly known as caustic effects.
In contrast to atom situation, where caustic effects are confined to a narrow energy regime around the HOHG cut-off energy and the enhancement due to caustic trajectory convergence is less apparent, the two-dimensional nature of graphene results in a broad energy region for HOHG enhancement that the caustic trajectories can even dominate the entire interband harmonic generation regime. The magnitude of enhancement is significant and can be estimated to be on the order of $\sim N^{2/3}$, with $N$ representing the harmonic order, according to the catastrophe theory.
These mechanisms have broad applicability and hold significant implications for other two-dimensional materials, as well as bulk materials, providing crucial insights into the understanding of HOHG phenomena in diverse material systems.

\end{abstract}
\affiliation{$^{1}$Graduate School, China Academy of Engineering Physics, Beijing 100193, China\\
$^{2}$Institute of Applied Physics and Computational Mathematics, Beijing 100088, China}

\maketitle

\textit{Introduction}.
As light propagates, it exhibits a fascinating phenomenon where multiple light rays converge, giving rise to bright focusing features known as caustic effects \cite{RThom,MBerry,YAKravtsov}.
In analogous to the situation of light propagation, the generalized caustic effects occur when multiple trajectories converge, resulting in the formation of a singularity and the subsequent enhancement of physical phenomena \cite{KWFord,PSikivie,MVBerry,QZXia,SBrennecke,LGLiao}.
The prediction of intensity enhancement can be achieved using catastrophe theory, which associates each caustic with a specific topological type of catastrophe \cite{YAKravtsov}.
These caustics can be observed in diverse fields, including acoustics \cite{FDTappert}, radio propagation \cite{TACroft}, as well as high-order harmonic generation (HOHG) \cite{ORaz,VABir,AJUzan,JChen}.

The generation of attosecond pulses \cite{Ferenc} has sparked considerable research interest in HOHG across various media, including gases \cite{ALHuillier,Corkum1,Lewenstein,Itatani,Meckel,YJChen}, crystalline solids \cite{RLu,Ghimire,Luu,Georges,RostCYu}, and two-dimensional materials \cite{Hanzhe,Rost,CHeide,Dong1,Dong2}.
When considering atoms irradiated by a femtosecond laser, caustic effects occur at the cut-off energy of HOHG. At this critical point, two branches of electron trajectories, commonly referred to as `short' and `long' trajectories, coalesce and contribute to the same harmonic energy, resulting in an enhancement in the spectrum magnitude \cite{ORaz,DFacciala}. Recently, the discussions of  caustic effects on HOHG are  extended to solid material such as MgO \cite{AJUzan}. It was claimed that  the Van Hove singularities in the energy band structure might result in the caustic singularity. However,  the enhanced HOHG spectra observed in the experiment apparently deviate from the locations of the Van Hove singularities \cite{AJUzan}.
On the other hand, theoretical investigations of the caustic effects on the HOHG of a one-dimensional periodic potential model has been made theoretically \cite{JChen}. Similar to atomic scenarios, it is found that the  caustic enhancement emerges only at a cut-off regime determined by the maximum electron-hole recombination energy \cite{JChen}.

In this Letter, we present the inaugural theoretical exploration of caustic effects in HOHG within the practical two-dimensional material, exemplified by the widely recognized material, graphene.
Graphene features a periodic hexagonal lattice with precisely two carbon atoms per unit cell.
We focus on investigating the caustic effects of HOHG in graphene under the influence of linearly polarized mid-infrared (MIR) laser irradiation. In contrast to atom scenario, we find that the two-dimensional nature of graphene leads to HOHG enhancement across a wide energy range, potentially encompassing the entire interband harmonics.
In particular, the location of HOHG enhancement peak is found to exactly  correspond to the zero determinant of  the Hessian matrix of the semiclassical action of the electron-hole recombination trajectories and  has nothing to do with the Van Hove singularities in the energy band structure.

%%%%%%%%%%%%%%%%%%%%%%%%%%%%%%%%%%%%%%%%%%%%%%%%%%%%%%%
\begin{figure}[t]
\begin{center}
\includegraphics[width=8.5cm,height=4.5cm]{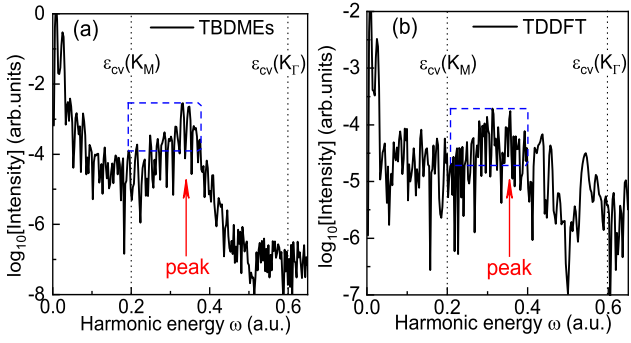}
\caption{
The harmonic spectra calculated by TBDMEs (a) and TDDFT (b) with the laser intensity of $8 \times 10^{11}$ W/cm$^{2}$ and the wavelength of $5500$ nm.
In panels (a) and (b), the obvious HES are marked by dashed rectangles, and the peaks of HES are indicated by red arrows.
The vertical dotted lines label the energy difference between the $c$ and $v$ bands at the Van Hove singularities ($M$ and $\Gamma$ points of graphene).
Notably, it is evident that the HES peaks are significantly far away from the Van Hove singularities.
}
\label{fig:graph1}
\end{center}
\end{figure}
%%%%%%%%%%%%%%%%%%%%%%%%%%%%%%%%%%%%%%%%%%%%%%%%%%%%%

\textit{Harmonic enhancement structure (HES).}
We perform calculations of HOHG using the two-band density-matrix equations (TBDMEs) as well as the  time-dependent density functional theory (TDDFT).
Here, the vector potential of the MIR laser field is $ \textit{\textbf{A}}(t) = \textit{A}_{0} \sin^2 (\omega_0 t / 2n) \sin(\omega_0 t) \textit{\textbf{e}}$, where $n = 3$, and $\textit{A}_{0}$ denotes the amplitude.
The frequency of the MIR laser field, denoted as $\omega_0$, corresponds to a wavelength of $\lambda = 5500$ nm.
The unit vector $\textit{\textbf{e}}$ indicates the direction along the $\Gamma - M$ axis of graphene \cite{SupplMate}.
Throughout the paper, atomic units are employed unless otherwise specified.
The harmonic spectra shown in Fig. 1 are simulated with the laser intensity of $8 \times 10^{11}$ W/cm$^{2}$.
It is apparent that the harmonic spectra obtained through both methods exhibit qualitative consistency.
In particular, the both spectra exhibit the apparent HES (marked by dashed rectangles) with a distinct  peak (marked by the red arrows). As comparison, we also label the Van Hove singularities ($M$ and $\Gamma$ points of graphene) with the vertical dotted lines in Fig. 1. Obviously, our calculated  HES peaks are  far away from the Van Hove singularities, similar to  the observations  in Ref. \cite{AJUzan}.

$\textit{Electron-hole recombination trajectory}$.
To unveil the underlying mechanisms behind the HES in graphene, we investigate the electron-hole recombination trajectories, leveraging the framework of TBDMEs \cite{SupplMate}.
Within the strong field approximation formulation \cite{Keldysh,GVampa}, the intraband currents become negligible and the interband currents play a dominant role in the  harmonic generation.
The Fourier transform of the total current can be expressed as:
\begin{align}
\textit{j}(\omega) \sim & \int_{\textrm{BZ}}  d \textrm{K}_{0x} \int_{\textrm{BZ}} d \textrm{K}_{0y} \int_{- \infty}^{\infty} d t \int_{-\infty}^{t} dt^{\prime} g(\textrm{K}_{0x},\textrm{K}_{0y},t,t^{\prime}) \nonumber\\
& \times e^{-iS(\textrm{K}_{0x},\textrm{K}_{0y},t,t^{\prime},\omega)} + c.c.,
\end{align}
where $\textbf{K}_{0} = (\textrm{K}_{0x},\textrm{K}_{0y})$ represents the lattice momentum within the first Brillouin zone ($\textrm{BZ}$).
$S(\textrm{K}_{0x},\textrm{K}_{0y},t,t^{\prime},\omega) = \int_{t^{\prime}}^{t}  \varepsilon_{c v}(\textrm{K}_{x}(\tau),\textrm{K}_{0y}) d \tau - \omega t$ denotes the semiclassical action, with $\textrm{K}_{x}(t) = \textrm{K}_{0x} + A(t)$.
The term $\varepsilon_{c v}(\textbf{k})$ represents the energy difference between the $c$ and $v$ bands for the lattice momentum $\textbf{k}$.
Notably, in contrast to the rapidly oscillating exponent, $g(\textrm{K}_{0x},\textrm{K}_{0y},t^{\prime},t)$ constitutes a slowly varying component within the expression \cite{SupplMate}.

%%%%%%%%%%%%%%%%%%%%%%%%%%%%%%%%%%%%%%%%%%%%%%%%%%%%%%%
\begin{figure}[t]
\begin{center}
\includegraphics[width=8.5cm,height=8cm]{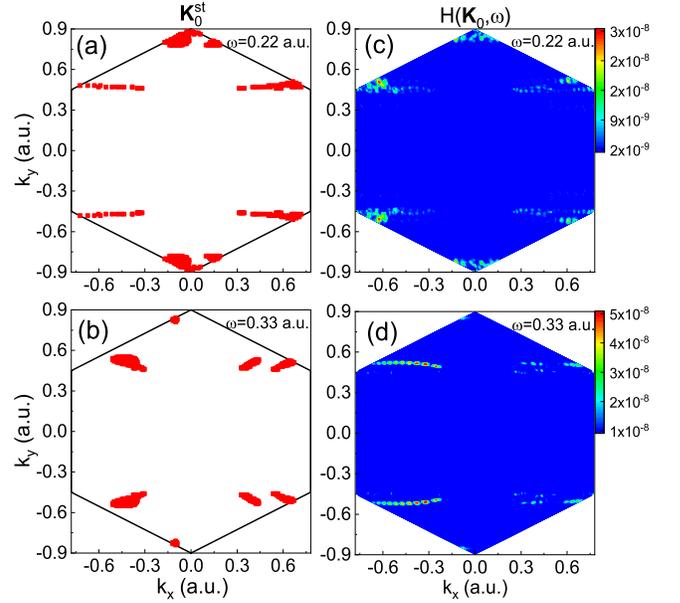}
\caption{
For the harmonic frequency of $\omega = 0.22$ a.u. [(a) and (c)] and $\omega = 0.33$ a.u. [(b) and (d)], the red dots in (a) and (b) mark the saddle point momenta $\textbf{K}_{0}^{st}$ calculated using Eqs. (2).
The harmonic intensities $H(\textbf{K}_0,\omega)$ calculated by TBDMEs are shown in (c) and (d).
}
\label{fig:graph2}
\end{center}
\end{figure}
%%%%%%%%%%%%%%%%%%%%%%%%%%%%%%%%%%%%%%%%%%%%%%%%%%%%%

In contrast to other solid materials such as MgO and ZnO \cite{AJUzan,GVampa}, graphene is unique lying in the existence of Dirac cones and zero energy gaps at the Dirac points.
Consequently, when employing the stationary phase approximation to all four integral variables in Eq. (1), a constraint arises where electrons in the valence band can only be excited to the conduction band through the Dirac points.
Based on our semiclassical trajectory calculations, under this constraint, the excited electrons do not have the opportunity to recombine with the hole \cite{YFeng}.
Our numerical simulations suggest that the  electrons excited  in the proximity of Dirac points can recombine with the hole and  emit harmonics successfully. With these considerations, we apply the stationary phase approximation only to three integral variables of $\textrm{K}_{0x}, \textrm{K}_{0y}, t$ and obtain following three saddle point equations,
\begin{subequations}
\begin{align}
\int_{t_i}^{t_r} \nabla_{\textrm{K}_{x}^{st}(\tau)} \varepsilon_{c v}\left(\textrm{K}_{x}^{st}(\tau), \textrm{K}_{0 y}^{st}\right) d \tau  = & 0,  \\
\int_{t_i}^{t_r} \nabla_{\textrm{K}_{0y}^{st}} \varepsilon_{c v}\left(\textrm{K}_{x}^{st}(\tau), \textrm{K}_{0 y}^{st}\right) d \tau  = & 0, \\
\varepsilon_{c v}\left(\textrm{K}_x^{st}\left(t_r\right), \textrm{K}_{0 y}^{st}\right)-\omega = & 0,
\end{align}
\end{subequations}
in which $t_{i}$ and $t_{r}$ represent the birth and recombination times of electron-hole pair, respectively.
$\textbf{K}_{0}^{st} = (\textrm{K}_{0x}^{st},\textrm{K}_{0y}^{st})$ is saddle point momentum and $\textrm{K}_{x}^{st}(t) = \textrm{K}_{0x}^{st} +  \textit{A}(t)$.
Equations (2a) and (2b) represent the conditions for the perfect electron-hole recombination trajectories, in contrast to the  imperfect recollisions \cite{LYue,AMParks,YFeng}.
The harmonic energy $\omega$ emitted during the  electron-hole pair recombination is given by Eq. (2c).

%%%%%%%%%%%%%%%%%%%%%%%%%%%%%%%%%%%%%%%%%%%%%%%%%%%%%%%
\begin{figure*}[t]
\begin{center}
\includegraphics[width=16cm,height=6.5cm]{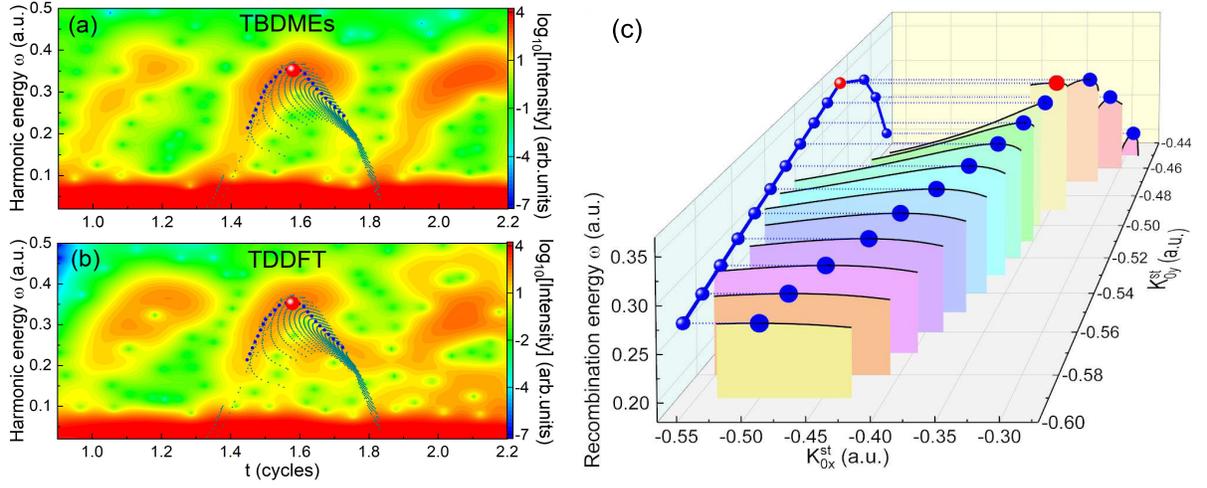}
\caption{
(a), (b) The time-frequency distributions corresponding to the harmonic spectra shown in Fig. 1.
In panels (a) and (b), the dark cyan points represent the recombination trajectories calculated by Eqs. (2).
(c) The recombination energy $\omega$ as a function of the saddle point momenta $\textbf{K}_{0}^{st} = (\textrm{K}_{0x}^{st},\textrm{K}_{0y}^{st})$.
The black curves show the relationship between the recombination energy $\omega$ and $\textrm{K}_{0x}^{st}$ for specific $\textrm{K}_{0y}^{st}$.
In panels (a), (b), and (c),
the blue dots indicate 1D caustic trajectories with $\partial \omega / \partial \textrm{K}_{0x}^{st} = 0$, and
the red points indicate 2D caustic trajectories with $\partial \omega /\partial \textrm{K}_{0x}^{st} = \partial \omega / \partial \textrm{K}_{0y}^{st} = 0$.
}
\label{fig:graph3}
\end{center}
\end{figure*}
%%%%%%%%%%%%%%%%%%%%%%%%%%%%%%%%%%%%%%%%%%%%%%%%%%%%%

To validate the applicability of our recombination trajectory theory for graphene, we examine the feasibility of our approach.
In Figs. 2(a) and 2(b), we illustrate the saddle point momenta $\textbf{K}_{0}^{st}$ calculated by Eqs. (2) as red dots, which correspond to recombination energies of $\omega = 0.22$ and $0.33$ a.u., respectively.
In Figs. 2(c) and 2(d), we present the harmonic intensities $H(\textbf{K}_0, \omega)$ at the frequency $\omega$, emitted by electrons with lattice momenta $\textbf{K}_0$  \cite{SupplMate}.
Upon comparing Fig. 2(a) with Fig. 2(c) and Fig. 2(b) with Fig. 2(d), it becomes evident that the saddle point momenta predicted by our recombination trajectory  align qualitatively with the lattice momenta that emit high-intensity harmonics as determined through numerical calculations.
These findings strongly suggest that our recombination trajectory theory is indeed applicable to graphene.

$\textit{Caustic effects on HOHG}$.
 The harmonic intensity, denoted by $H(\omega)$, can be evaluated using $H(\omega) = \omega^2 \vert \textit{j}(\omega) \vert^{2}$, where $\textit{j}(\omega)$ from Eq. (1) can be deduced  according to the saddle point trajectories that satisfy Eqs. (2):
\begin{align}
\textit{j}(\omega) \sim & \sum_{\textrm{K}_{0 x}^{s t}, \textrm{K}_{0 y}^{s t}, t_r, t_i} g\left(\textrm{K}_{0 x}^{s t}, \textrm{K}_{0 y}^{s t}, t_{r}, t_i\right) \times \nonumber\\
& \frac{e^{-i S\left(\textrm{K}_{0 x}^{s t}, \textrm{K}_{0 y}^{s t}, t_r, t_i,\omega\right) }}{\sqrt{\left|\operatorname{det}\left[S^{\prime \prime}\left(\textrm{K}_{0 x}^{s t}, \textrm{K}_{0 y}^{s t}, t_r, t_i, \omega \right)\right]\right|}} + c.c.
\end{align}
Here, $S^{\prime \prime}(\textrm{K}_{0x}^{st},\textrm{K}_{0y}^{st},t_r,t_i,\omega)$ is the Hessian matrix of the semiclassical action $S(\textrm{K}_{0x}^{st},\textrm{K}_{0y}^{st},t_r,t_i,\omega)$ with respect to $\textrm{K}_{0x}^{st}$, $\textrm{K}_{0y}^{st}$ and $t_r$, whose determinant is
\begin{align}
\operatorname{det}\left[S^{\prime \prime}\right]=\frac{\partial \omega}{\partial \textrm{K}_{0 x}^{st}} \mathcal{H}_1-\frac{\partial \omega}{\partial \textrm{K}_{0 y}^{st}} \mathcal{H}_2-E\left(t_r\right) \frac{\partial \omega}{\partial \textrm{K}_{0 x}^{st}} \mathcal{H}_3.
\end{align}
Here, $\mathcal{H}_1$, $\mathcal{H}_2$ and $\mathcal{H}_3$ are the second order determinants \cite{SupplMate}.

We can then obtain following caustic equations,
\begin{align}
\partial \omega / \partial \textrm{K}_{0x}^{st} =0,\,\,\, \partial \omega / \partial \textrm{K}_{0y}^{st} = 0,
\end{align}
which determines a specific saddle point trajectory that originates from the  lattice momenta of $(\textrm{K}_{0x}^{st*},\textrm{K}_{0y}^{st*})$ and finally emits a harmonic photon with the energy $\omega^*$ through a perfect electron-hole recombination.
It also implies that the saddle point trajectories originating from the vicinity of $(\textrm{K}_{0x}^{st*},\textrm{K}_{0y}^{st*})$  tend to emit harmonic photons with the same energy $\omega^*$, demonstrating a kind of 2D trajectory caustic phenomenon.
According to Eqs. (3) and (4), one can find that, for this specific trajectory, the determinant of the Hessian matrix $S^{\prime \prime}(\textrm{K}_{0x}^{st*},\textrm{K}_{0y}^{st*},t_r^*,t_i^*,\omega^*)$ becomes zero and the corresponding harmonic intensity diverges into infinity.
This caustic singularity indicates a potentially significant amplification in the magnitude of the harmonics in the energy range around $\omega^*$.

Using the field parameters in Fig. 1 and with the help of saddle point equations (2), we have solved the caustic equations (5) and obtained  $\omega^*=0.35$ a.u., which is in complete agreement with the location of the HES peaks illustrated in Fig. 1.
Notice the caustic singularity is totally different from the Van Hove singularities \cite{AJUzan} of energy bands that are determined by $\vert \nabla_{\textbf{k}} \varepsilon_{cv}(\textbf{k})\vert = 0$ and correspond to $\omega=0.2$ and $0.6$ a.u. as shown in Fig. 1.

In Figs. 3(a) and 3(b), we present the time-frequency distributions corresponding to the harmonic spectra depicted in Fig. 1 \cite{SupplMate}.
The results obtained from both TBDMEs and TDDFT demonstrate qualitative agreement.
The red points correspond to the specific 2D caustic trajectory.
One can find that the red point exactly locates at  the brightest spot of the time-frequency distributions.

Corresponding to the recombination trajectories shown in Figs. 3(a) and 3(b), we illustrate the recombination energy $\omega$ as a function of the saddle point momenta $\textbf{K}_0^{st} = (\textrm{K}_{0x}^{st},\textrm{K}_{0y}^{st})$ in Fig. 3(c).
In one-dimensional sections for a fixed  $\textrm{K}_{0y}^{st}$, the blue dots indicate the local  maxima of the recombination energies where $\partial \omega / \partial \textrm{K}_{0x}^{st} = 0$.
These trajectories are also marked by the blue dots in Figs. 3(a) and 3(b).

The blue dots in Figs. 3(a) and 3(b) represent trajectories that only  satisfy $\partial \omega / \partial \textrm{K}_{0x}^{st} = 0$ for a fixed  $\textrm{K}_{0y}^{st}$, and are referred to as 1D caustic trajectories.
Additional calculations reveal that for these trajectories, the determinant of $S^{\prime \prime}(\textrm{K}_{0x}^{st},\textrm{K}_{0y}^{st},t_r,t_i,\omega)$ is not zero but is relatively small, indicating a relatively higher harmonic enhancement.
It is noteworthy that the blue dots in Figs. 3(a) and 3(b) are approximately situated at the central area of the highlighted time-frequency distributions, suggesting that these particular trajectories may play a dominant role in the generation of interband harmonics \cite{SupplMate}.

%%%%%%%%%%%%%%%%%%%%%%%%%%%%%%%%%%%%%%%%%%%%%%%%%%%%%%%
\begin{figure}[t]
\begin{center}
\includegraphics[width=8.5cm,height=7.5cm]{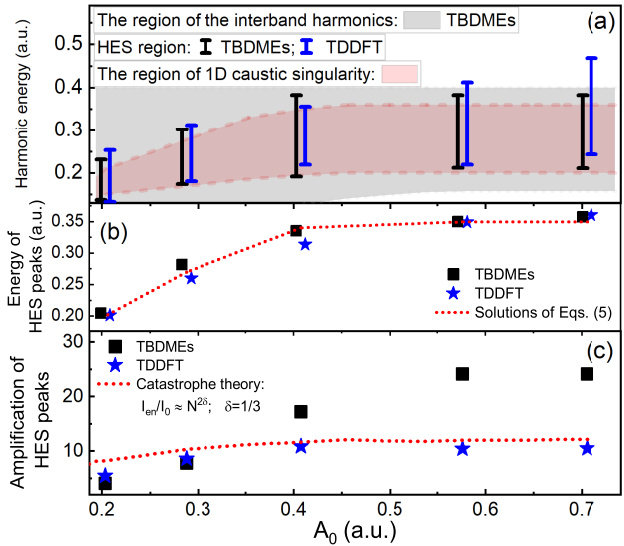}
\caption{(a) The gray area corresponds to the region where the interband currents dominate the harmonic generation according to  TBDMEs \cite{SupplMate}.
The energy regions of  HES are indicated by bars (e.g. refer to the dashed rectangles in Fig. 1).
The pink area indicates the energy regions where 1D caustic singularities emerge.
(b) The black squares and blue stars indicate the  energies corresponding to the  HES peaks, and the red lines are the predictions of caustic equations (5).
(c) The black square and blue star show the amplification of harmonic intensity at the HES peaks. The red line is the  prediction of the catastrophe theory.
}
\label{fig:graph5}
\end{center}
\end{figure}
%%%%%%%%%%%%%%%%%%%%%%%%%%%%%%%%%%%%%%%%%%%%%%%%%%%%%

$\textit{Laser parameter dependent caustic effects}$.
We perform extensive calculations of HOHG across a broad range of laser intensities.
The HES information as a function of amplitude $A_{0}$ of the laser vector potential is shown in Fig. 4.
Figure 4(a) shows that the regimes of 1D caustic singularity are qualitatively consistent with  the energy region of HES simulated by both TBDMEs and TDDFT. In contrast to  atom situation where caustic effects are limited to a narrow  regime  around the cut-off energy of HOHG \cite{ORaz,SupplMate},  the caustics in graphene will lead to a broad energy region of HOHG enhancement and  even dominate the entire interband harmonic generation process.
Figure 4(b) clearly demonstrates that  the locations of HES peaks can be well predicted by  the caustic equations (5).

The relative enhancement of the HES peaks can be evaluated by the catastrophe theory $I_{en} / I_{0} \approx N^{2 \delta}$ \cite{YAKravtsov,ORaz}.
Here, $I_{en}$ represents the intensity at the caustic peaks, and $I_{0}$ is the intensity far from the caustic region.
$N$ is the harmonic order corresponding to  the caustic peak.
The focusing index $\delta$ depends on the types of catastrophes, which are determined by the number of the control parameters and state variable.
In the case of atoms excited by linearly polarized monochromatic laser field, the harmonic amplitude can be evaluated by
$E(\omega) = \int E_{\textrm{XUV}}(t_i, \omega) e^{-iS_{0}(t_i,\omega)} d t_i$ \cite{ORaz}, in which $E_{\textrm{XUV}}(t_i, \omega)$ is the amplitude of the quantum trajectory associated with the ionization time $t_i$.
In the semiclassical action $S_{0}(t_i,\omega)$, there is only one control parameter ($\omega$) and one state variable ($t_i$), corresponding to the fold catastrophe with $\delta = 1/6$.
To amplify  the caustic effect, in Ref. \cite{ORaz}, the authors increase the number of control parameters by using a two-colour laser. Then, the type of catastrophes turns to be  swallowtail, corresponding $\delta = 3/10$.

In our case of graphene irradiated by a linearly polarized MIR laser field, according to the saddle point equations (2),  there are  two state variables of $\textrm{K}_{0x}^{st}$, and $\textrm{K}_{0y}^{st}$.
Then, the harmonic amplitude can be evaluated as
$E(\omega) = \int \int d \textrm{K}_{0x}^{st} d \textrm{K}_{0y}^{st} g(\textrm{K}_{0x}^{st}, \textrm{K}_{0y}^{st}, \omega) e^{-iS(\textrm{K}_{0x}^{st}, \textrm{K}_{0y}^{st},\omega)}$.
The types of catastrophes turn to be elliptic umbilic or  hyperbolic umbilic with
the focusing index $\delta=1/3$.  If we assume that $\textrm{K}_{0y}^{st}$ is fixed, the above 2D caustic singularity will degenerate to be 1D caustic singularity corresponding to the fold catastrophe with focusing index $\delta = 1/6$, in analogous to atomic or 1D periodic potential cases \cite{ORaz,JChen}.
Figure 4(c) shows that our numerical results are qualitatively consistent with the predictions of the catastrophe theory.

\textit{Summary.}
Our numerical simulations with both TBDMEs and TDDFT uncover
a striking HES for the HOHG in graphene, which we attribute to the fantastic caustic effects.
We have developed the electron-hole recombination trajectory theory and then deduced the caustic equations that can precisely predict the location of the HES peak as well as width of HES.
With the help of the catastrophe theory, we can also estimate the magnitude of the enhancement in graphene's HOHG.
Our findings can be experimentally observed utilizing contemporary techniques \cite{NYoshikawa}, and our theoretical analysis holds relevance for other two-dimensional materials as well as bulk materials.

\section*{ACKNOWLEDGMENTS}

This work is supported by NSAF (Grant No. U1930403)
and the National Natural Science Foundation of China (Grants No. 11974057).
We acknowledge valuable discussions with Dr. Jiaxiang Chen.

\end{document}